%% file: main.tex
\documentclass[vecphys]{svmult}

\usepackage{makeidx}         
\usepackage{graphicx}        
\usepackage{multicol}        
\usepackage[bottom]{footmisc}

\makeindex             

\begin{document}

\title*{Classical and quantum breakdown in disordered materials}
\author{Debashis Samanta\inst{1}, Bikas K. Chakrabarti\inst{2}\and
Purusattam Ray\inst{3}}
\institute{Theoretical Condensed Matter Physics Division, 
Saha Institute of Nuclear Physics,
1/AF Bidhannagar, Kolkata 
700064, India
\texttt{debashis.samanta@saha.ac.in}
\and Theoretical Condensed Matter Physics Division and Centre for Applied 
Mathematics and Computational Science, Saha Institute of Nuclear Physics,
1/AF Bidhannagar, Kolkata 700064, India
\newline
\texttt{bikask.chakrabarti@saha.ac.in}
\and The Institute of Mathematical Sciences,
C.I.T. Campus, Taramani, Chennai 600113, India
\newline
\texttt{ray@imsc.res.in}}
%
%
\maketitle


\section{Introduction}
\input{intro}
\section{Analysis of the fuse problem}
\subsection{Disordered fuse network}

\input{fuse}
\subsection{Distribution of the failure current}
\input{bre}

\subsection{Continuum model}
\input{continuum}
\subsection{Electromigration}
\input{mig}
\subsection{Numerical simulations of random fuse network}
\input{fusen}
\section{Dielectric breakdown problem}
\input{di}
\input{tdi}

\input{sto}
\section{Zener breakdown in Anderson insulators}
\input{qua2}
\section{Conclusions}
\input{conclu.tex}

\input{bbl2}                     
\printindex
\end{document}

%% file: intro.tex

It is our common household experience that when the voltage drop across 
a fuse exceeds a limit, the fuse burns out.  A fuse is nothing but 
a conductor which conducts uniform current under an applied voltage 
upto a certain limit beyond which it burns out and becomes nonconducting.
 This is called fuse failure\index{fuse failure}. Similarly in a dielectric breakdown, 
a dielectric starts to conduct 
electricity when the voltage drop across it attains certain threshold value.
The above two phenomena are examples of breakdown process  which is described broadly 
as the failure of a physical attribute when the perturbing force driving it goes beyond 
a limiting value. The most common example of the process is the breaking of a material at a high 
stress beyond its strength.

Naturally occuring solids are almost always inhomogeneous 
and have defects like vacancies, microcracks or impurities.
These defects are weak points across which stress fields or 
electric fields or current densities concentrate. 
Depending on the geometry of the defect, 
the concentrated field can be very high. If the field or density
exceeds
the material fatigue limit, failure nucleates locally around the 
defect and starts to propagate. The propagation can be arrested, or can 
spread to the entire system leading
to the global failure of the system, depending on the field strength 
and the defect structure in the material. The failure of a material, 
thus, depends largely on the disorder present in it.

The simplest form of the weak points is substitutional disorder which 
can be realised by the inclusion of 
nonconducting material in a conducting material or vice versa. 
Since we are interested in the macroscopic properties of global failure,  
we consider failures at length scale 
larger than the regions in which the local failure appears (i.e., 
atomic distances). At this scale, the disordered fuse system can be modelled  
by a lattice whose bonds are conducting with a certain probability $p$ 
and nonconducting otherwise.

For such models, one can apply the principles of percolation theory 
\cite{Stauffer and Aharony:1992_SCR}. Below the percolation 
threshold\index{threshold!percolation} $p_c$
(for $p<p_c$) the system is not connected globally by channels of conducting bonds and 
so nonconducting, whereas for $p>p_c$ the 
system is conducting (at least one continuous path exists across 
the system via the conducting bonds). 

\begin{figure}[htb]
\centering
      \includegraphics[height=4cm]{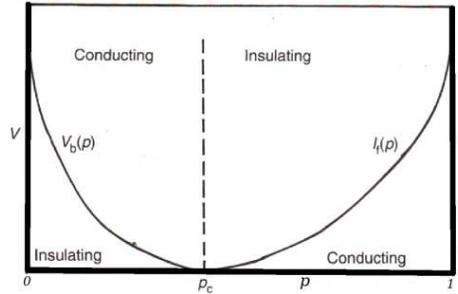}
      \caption{\footnotesize{Phase diagram of a mixture 
                containing $p$ fraction of conductors
                and $(1-p)$ fraction of insulators at random.
                On the left side, the sample
                is insulating for voltage $V$ less than the breakdown voltage
                $V_b(p)$ and conducting otherwise.
                On the right side, the system is conducting
                for current $I$ less than the fuse current $I_f(p)$ and 
                insulating otherwise.}}    
      \label{fig:fpscan_SCR}
\end{figure}

For $p>p_c$, when conducting channels span across the system and the system 
is conducting, one can ask what is the fuse current\index{fuse failure!current} 
$I_f$ required 
so that the system becomes nonconducting. At $I>I_f (p)$ not a single spanning 
path of conducting bonds exists. At $p=1$ all the conducting bonds are present
and $I_f(1)$ normalized by the sample size is simply the fuse 
threshold current\index{threshold!fuse current}
for each of the conducting bonds.
For $p<p_c$, the conducting bonds do not form a continuous path across
the sample, and the sample is insulating. With increasing voltage $V$ across the sample,
one can get a continuous path through original conductors and broken 
dielectric for $V \geq V_b(p)$, where $V_b(p)$ is the 
dielectric breakdown voltage\index{dielectric breakdown!voltage}. 
At $p=0$, all the non
conducting bonds are there in the system and $V_b(0)$ normalized by the 
system size is simply the breakdown threshold\index{threshold!dielectric breakdown voltage}
 of each of the nonconducting
bonds.   
On the other hand, as 
$p \rightarrow p_c$, both the dielectric breakdown 
voltage\index{dielectric breakdown!voltage} and fuse failure 
current\index{fuse failure!current} tend to zero. 
The corresponding phase diagram 
is shown in fig.~\ref{fig:fpscan_SCR}. 
 
In this article, we will review the basic ideas in breakdown problems in 
terms of the fuse failure and dielectric breakdown problems and then discuss 
how these ideas can be extended to the breakdown in quantum mechanical 
systems such as Anderson insulators. In disordered electronic systems above
two dimensions, the electronic states below the mobility edge are all 
localised (the system with Fermi level within this range behaves as 
an insulator)
and the states above the mobility 
edge are extended (the system turns to a conductor for Fermi level in this
range). The Anderson transition\index{Anderson transition} 
from insulating to conducting phase across the mobility edge is well 
studied \cite{Lee and Ramakrishnan:1985_SCR}. We discuss here the possibility 
of a breakdown from insulating to conducting phase by applying 
strong electric field and compare the quantum breakdown with the classical 
dielectric or fuse breakdown in disordered materials.


%% file: fuse.tex
In a pure conductor placed between two electrodes with a potential
difference, the field lines within the conductor are all parallal to each 
other and perpendicular to the electrode surfaces. In presence of a disorder 
in the form of an insulating region, the field lines get deformed around the 
defect. As a result in the vicinity of the defect current density increases 
to $i_e$ from $i$, the current density value far away from the defect. 
So one can write 

\begin{equation}
i_e=i(1+k)
\end{equation}
where $k$ is the enhancement factor which depends on the geometry of the 
defect. As an example, $k$ becomes $l/b$ for an elliptic defect 
of semi-major axis $l$ and semi-minor axis $b$ \cite{Chakrabarti:1997_SCR}.
For electrode surface area $S$ total current $I$ is 

\begin{equation}
I=Si=\frac{Si_e}{1+k}.
\end{equation}

Failure occurs for the first time when $i_e$ becomes equal to $i_0$ which is 
the fatigue limit of the sample material.
The failure current is then 
\begin{equation}
I_f=\frac{Si_0}{1+k}.
\end {equation}

Larger the enhancement factor $k$, smaller is the failure current $I_f$.
For an example, if $l>>b$ for an elliptic defect, $I_f$ may get reduced 
by a large extent. So the presence of defects in the material facilitates 
failure and the presence of sharper edges of defects make the system
 more vulnerable to failure. The 
failure makes the defect bigger and hence $k$ larger. This means that the 
current density around the defect enhances further causing failure again.
The process causes rapid failure of the whole sample. This means that 
external voltage for the global failure is the same as the voltage for 
the first 
local failure and a local failures once started leads to the failure of the 
entire sample. This type of failures are 
called \textit{brittle failures}\index{brittle failure}.

So far we have discussed the influence of a single defect of a regular size 
within the sample. Natural and engineering samples usually contain a large 
number of defects of irregular shapes and sizes at random. To get some 
quantitative estimate of the failure criteria in terms of defect parameters 
we idealize the solid to lattice model and use percolation theory  for defects
\cite{Ray and Chakrabarti:1985_SCR, Ray:1985_SCR, Duxbury et al:1986_SCR}. 

We start with a hypercubic lattice in $2d$ or $3d$ 
with all conducting bonds. 
The simplest defect can be introduced by removing one bond 
parallal to the direction of the current flow.
In this case, failure current is calculated as
\begin{equation}
I_f=\frac{\pi}{4}Li_0  
\end{equation}
in $2d$, where $Li_0$ is the fracture current of the lattice in absence of 
any defects and $L$ is the linear size of the lattice. The 
enhancement factor is $4/\pi$ here. We can introduce randomness in shape and 
size of the defect by removing $(1-p)$ fraction of the bonds randomly. It is 
no longer possible to determine the most vulnerable defect and to calculate the 
enhancement factor. We 
consider two limits: (1) dilute limit $(p \rightarrow 1)$ when the defect 
density is small and (2) near the critical point $p_c$ where the defect 
density is large and beyond which the lattice loses its connectivity.

\subsubsection{Dilute limit $(p\to 1)$}

In this limit there are only a few isolated defects (insulating bonds) in the 
sample. The current density around a defect is effectively 
independent of the presence 
of other defects. Under such considerations the most vulnerable of the defects 
will lead to the failure of the system. So our primary task is to identify 
the \textit{most probable dangerous defect} (the weakest point which 
causes the largest 
concentration of the current density).

The ensemble of $n$ successively removed bonds, far from 
boundaries \cite{Duxbury et al:1986_SCR, Duxbury et al:1987_SCR}, 
in a plane perpendicular to the current 
flow, act as dangerous defect. In $2d$ it is a linear defect and in $3d$  
the defect has the shape of a shape of disc. The current through the 
conducting bonds at the immediate neighbourhood  of the defect is
\begin{equation}
\label{eqn:imn_SCR}
i_e=i(1+k_2n) \hspace{5mm}(\mbox{in} \hspace{1mm} 2d),
 \hspace{20mm} i_e=i(1+k_3n^{\frac{1}{2}}) \hspace{5mm} (\mbox{in} \hspace{1mm} 3d).
\end{equation}
Here $i$ is current density through the bonds far away from the defect.
Enhancement factor $k$ contains $\sqrt{n}$ term because current is 
diverted by $n$ defects to spread uniformly around the perimeter 
of the disc, which is proportional to $\sqrt{n}$. 
The probability of appearance of a defect of $n$ successive insulating 
bonds is 
\begin{equation}
P(n) \sim (1-p)^nL^d,
\end{equation} 
where $L$ is the lattice size and $d$ is the dimension 
of the lattice. $L^d$ provides the number of places that 
the defects can occupy. The approach of $P(n) \approx 1$ gives the size of 
most probable dangerous defect:
\begin{equation}
\label{eqn:nc_SCR}
n_c=-\frac{2}{\ln (1-p)}\ln L \hspace{5mm} 
(\mbox{in} \hspace{1mm} 2d), 
\hspace{10mm}
n_c=-\frac{3}{\ln (1-p)}\ln L \hspace{5mm} 
(\mbox{in} \hspace{1mm} 3d).
\end{equation} 
Combining (\ref{eqn:imn_SCR}) and (\ref{eqn:nc_SCR}) $i_e$ becomes
\begin{eqnarray}
\label{eqn:im_SCR}
i_e & = & i\left[1+k_2\left(-\frac{2~\ln L}{\ln (1-p)}\right)\right]
\hspace{5mm} (\mbox{in} \hspace{1mm} 2d), \nonumber \\
    & = & i\left[1+k_3\left(-\frac{3~\ln L}{\ln (1-p)}\right)^{\frac{1}{2}}\right]
\hspace{2mm} (\mbox{in} \hspace{1mm} 3d).
\end{eqnarray}
Here, the total current is $iL^{(d-1)}$.
Equating $i_e$ with the threshold\index{threshold!fuse current} value $i_0$, 
the expression for the failure current\index{fuse failure!current} becomes
\begin{eqnarray}
\label{eqn:ifp1_SCR}
I_f=\frac{i_0L}{1+2k_2[\frac{\ln L}{\ln (1-p)}]}
\hspace{5mm}(\mbox{in} \hspace{1mm} 2d),
\nonumber
\\
I_f=\frac{i_0L^2}{1+\sqrt{3}k_3[\frac{\ln L}{\ln (1-p)}]^{\frac{1}{2}}}
\hspace{2mm}(\mbox{in} \hspace{1mm} 3d).
\end{eqnarray}

From the equations it is clear that as $p\to1$, $I_f$ reduces 
to $i_0L^{d-1}$, the value of the current in the lattice with all the bonds conducting. 
The slope of the $I_f(p)$ vs. $p$ curve at $p=1$ is infinite 
(see fig.~\ref{fig:fpscan_SCR}). It is expected because even the presence of 
a single defect enhances the process of breakdown through cascade. The most 
important thing is the size dependence of $I_f$. For large enough $L$ and $p$ 
not very close to 1 (such that the absolute value of $\ln (1-p)$ is not too 
large) the failure current per bond $i_f$($=I_f/L^{d-1}$) decreases 
as $1/\ln L$ and $1/(\ln L)^{1/2}$ in $2d$ and $3d$ respectively.

\subsubsection{Critical behavior ($p \to p_c$)}
\label{sec:criticality_SCR}

\begin{figure}[htbp]
   \begin{center}
      \includegraphics[width=0.49\textwidth]{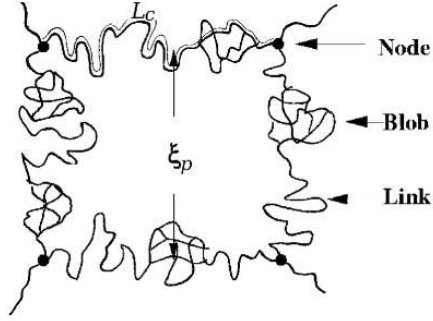}\hspace{2mm}
      \caption{\footnotesize{A portion of the node-link-blob 
              superlattice model near
              $p_c$. The distance between two nodes of the lattice is
              $\xi_p$, while chemical length\index{chemical length} of the 
              tortuous link of the
              super lattice is $L_c$.}}
      \label{fig:node_scr}
   \end{center}
\end{figure}

Near criticality $(p \to p_c)$, the material is strongly disordered and 
can be described by \textit{node-link-blob} 
picture\index{node-link-blob model} 
\cite{Stanley:1977_SCR,Skal and Shklovskii:1975_SCR,de Gennes:1976_SCR} 
of percolation theory (see fig.~\ref{fig:node_scr}). 
Close to $p_c$ the conducting part of the material 
extends over the sample to form percolating cluster which 
is \textit{self-similar} up to the length scale $\xi_p$, the percolation 
correlation length\index{correlation length}. The geometry of the cluster at 
length scale $\xi_p$ appears same as that of the original cluster at 
smaller scale.
 This means that the infinite cluster may be divided into cells of size $\xi_p$.
Each cell consists of backbone bonds (which takes part in 
current conduction) and numerous dangling bonds (bonds which do not take part 
in current conduction). Only backbone is important because it takes part in 
conductivity. The backbone is made of two kinds of bonds: multiply 
connected bonds appear as blobs; and singly connected 
bonds (called \textit{red} bonds), appear as links. Distribution of 
current in the sample is solely determined by links; 
being singly connected, each link has to carry the full current inside a cell. 
Since 
at threshold $(p_c)$ correlation length $(\xi_p)$\index{correlation length} 
spans 
the sample (large enough), it is reasonable to assume that 
the failure current approaches zero as $p \to p_c$.

To determine quantitatively the critical behavior of the failure 
current\index{fuse failure!current} $I_f$,  
we consider the voltage $V$ applied across the node-link-blob 
network. Since $V$ is distributed among $L/\xi_p$ 
number of links (or cells) in series, the average voltage across each 
link is $V_L \sim \xi_p V$. The resistance of the sample $R$ is related 
to average link resistance $R_L$ through $R \sim \xi_p^{d-2}R_L$, since
there are $L \xi_p^{-1}$ number of links in series in the length $L$ of 
the sample and there are $\xi_p^{-1}$ number of parallal such links. So the 
mean current in a link appears to be $i_L \sim \xi_p^{d-1} V/R$ through the 
relation $i_L=V_L/R_L$. $V$ equals to failure or fuse voltage $V_f$ when 
$i_L$ reaches the maximum current which a link can stand. One can assume fuse 
voltage behaves as 
\begin{equation}
V_f \sim (p-p_c)^{-t_f}.
\end{equation}
Using the relation $\xi_p \sim (p-p_c)^{-\nu}$ and $R \sim (p-p_c)^{-t_c}$, $t_f$
comes out as $t_f=t_c-(d-1)\nu$.
From the relation $I_f=V_f/R$, we get

\begin{equation}
\label{eqn:ifp_SCR}
I_f \sim (p-p_c)^{(d-1)\nu}.
\end{equation}  

The values of the correlation length 
exponent\index{exponent!correlation length} $\nu$ are $1.33$ 
in $d=2$ and $0.88$ in $d=3$ and the values of the 
conductivity exponent\index{exponent!conductivity in discrete model} $t_c$ are $1.33$ 
in $d=2$ and $2$ in $d=3$ \cite{Stauffer and Aharony:1992_SCR}. 
So it is clear that fuse voltage\index{exponent!fuse failure voltage in discrete model} 
$V_f$ attains 
finite value in $2d$ and diverges with the exponent $0.2$ in $3d$, in contrary 
to the failure current\index{exponent!fuse failure current in discrete model}
 which always approaches zero as $p \to p_c$.

\subsubsection{Influence of the sample size}

Size dependence of $I_f$ is related to the notion of the most dangerous 
defect present in the sample. In the present context the most dangerous 
defect is a `cell' of the infinite cluster with length $\xi_p$ in the 
direction parallal to the applied voltage and $l_{max}$ in the 
perpendicular direction. The total probability of having a defect of 
size $l$ is, $P=g(l)(L/\xi_p)^d$, where $g(l)$ is the probability 
density of defect cluster of linear size $l$. Percolation 
theory predicts \cite{Stauffer and Aharony:1992_SCR,Sahimi:1994_SCR} that
\begin{equation}
P \sim exp \left ( - \frac{l}{\xi_p}\right) \left ( \frac{L}{ \xi_p }\right)^d.
\end{equation}
$l_{max}$ is obtained when $P \approx 1 $ and
\begin{equation}
l_{max} \sim \xi_p \ln L.
\end{equation}  
Now the current that flows through the side link of the defect is 
proportional to $(l_{max})^{d-1}I$ and one obtains 
\begin{equation}
\label{eqn:ifsize_SCR}
I_f \sim \frac{(p-p_c)^{(d-1) \nu }}{(\ln L/ \xi_p )^{(d-1)}}.
\end{equation}
So this is the correction over equation~(\ref{eqn:ifp_SCR}) due 
to finite size of 
the sample. Bergman and Stroud \cite{Bergman and Stroud:1992_SCR} 
gave an idea about the competition 
between extreme statistics and percolation statistics. The extreme statistics
(size dependence of the most probable defect and 
failure\index{fuse failure!current} current) is 
expected to dominate for $\ln L> \xi_p $ (or, when 
Lifshitz scale\index{Lifshitz scale} is greater 
than the connectedness correlation length).  The dominance of extreme 
statistics is expected for $p$ far away from percolation threshold when the 
correlation length $\xi_p$\index{correlation length} is small. 

Li and Duxbury \cite{Li and Duxbury:1987_SCR} proposed the dependence 
of $I_f$ on $L$ through $(\ln L)^{- \psi_f}$. 
The approximate range 
of $\psi_f$\index{exponent!size dependence of fuse failure current} is 
\begin{equation}
\label{eqn:rangepsi_SCR}
\frac{1}{2(d-1)} < \psi_f < 1.
\end{equation}
Considering the results of the dilute limit and in the critical region, the combined
form of~(\ref{eqn:ifp1_SCR}) and~(\ref{eqn:ifp_SCR}) is

\begin{equation}
\label{eqn:unifiedif_SCR}
I_f=I_0 \frac{\left [ \frac{(p-p_c)}{(1-p_c)}\right]^{\phi_f}}{1+K \left [- \frac{\ln (L/ \xi_p)}{\ln (1-p)}\right ]^{\psi_f}}.
\end{equation}
The value of the different 
exponents\index{exponent!fuse failure current in discrete model} 
(see table~\ref{tab:fuseexponent_SCR}) 
and 
the constant $K$ 
depend on the dimension and on the type of percolation. 
From~(\ref{eqn:ifp1_SCR}) 
and~(\ref{eqn:ifsize_SCR}) one can see that the combined result 
is valid only for $2d$.
 The expression has three obvious features:
\begin{enumerate}
\item
For $p=1$, $I_f=I_0$, as expected.
\item
Near $p=1$, $(p-p_c)$ is almost constant and we get back the 
expression~(\ref{eqn:ifp1_SCR}).
\item
Near $p_c$, the denominator of (\ref{eqn:unifiedif_SCR}) is of the order 
of unity and we recover the expresion~(\ref{eqn:ifp_SCR}) 
with $\phi_f=(d-1)\nu$.  
\end{enumerate}


%% file: bre.tex
\subsubsection{dilute limit ($p \to 1$)}

In random fuse networks the failure current\index{fuse failure!current} 
$I_f$ shows large 
sample to sample fluctuations. Since the failure current is determined 
by the weakest defect in the sample, the fluctuations in the failure 
currents do not come down with the system size. The average failure 
current is not a self-averaging quantity.

Distribution of the failure currents\index{fuse failure!current} in a 
system of size $L$  
follows as (see ref. \cite{Chakrabarti:1997_SCR}) 
\begin{equation}
\label{eqn:fibk1_SCR}
F_L(I) = 1-\exp\left\lbrack -A_d L^d \exp\left \lbrace - dA \ln L \left( \frac{ \frac{I_0}{I}-1}{ \frac{I_0}{I_f}-1}\right)^{d-1}\right \rbrace \right\rbrack.
\end{equation}

The derivative of this cumulative failure probability distribution 
$F_L(I)$ with respect  to current $I$ provides the current $I$ at 
which the system fails. Certainly, at most probable failure current the 
derivative becomes maximum. Here $I_0$ is the failure 
current\index{fuse failure!current} for pure sample. 
It may become obvious with a simple  
calculation that $I_f$ appears to be the  most probable 
failure current (as was assumed in calculation) only in the limit of 
large enough system size.

Though the current $I$ can vary from zero to infinity, the form of $F_L(I)$
is meaningful only for $I$ upto $I_0$.  
One should expect the value unity for $F_L(I)$ 
at $I=I_0$ and $I=\infty$, but it is true only for large system size.
$F_L(I)$  suffers from size and defect concentration dependence through $I_f$
 (see equation~\ref{eqn:ifp1_SCR}).

The above expression is reffered to as the
 Gumbel distribution\index{distribution function!Gumbel}
 \cite{Gumbel:1958_SCR}. Another well known one which
 is very often used in engineering is 
Weibull distribution\index{distribution function!Weibull}
\begin{equation}
F_L(I)=1-\exp \left \lbrack -rL^d \left(\frac{I}{I_f} \right)^m\ \right \rbrack.
\end{equation}
Here $m$ is a constant and for large $m$ (say more than 5) $I_f$ refers 
to most probable failure current.

\subsubsection{At critical region $(p \to p_c)$}

Near criticality the cumulative failure distribution function is
\begin{equation}
\label{eqn:fipc_SCR}
F_L(I)=1-\exp \left \lbrack - A^\prime_dL^d \exp \left (- \frac{k^\prime(p-p_c)^\nu }{I^{\frac{1}{d-1}}}\right ) \right \rbrack,
\end{equation}
where $A^\prime_d$ and $k^\prime$ are two constants. 

%% file: continuum.tex
\label{sec:continuum_SCR}
One can extend the ideas of lattice percolation to continuum conducting medium. 
A material at the scale of the size of the defects can be looked upon as
a continuous field with some defects as inclusions in the field.  In continuum 
model of lattice percolation, insulating spherical holes (circular holes 
in $2d$) are punched at random as defects in a uniform conducting sample. 
The holes can overlap 
(\textit{Swiss-cheese model}) 
and two non-overlapping neighbouring
holes have a conducting region between them. These regions constitute 
conducting channels of cross-section $\delta$. Somewhat similar to the 
breakdown problem discussed above, the transport properties of any such 
channel depend on the transport capacities (cross-section and length) of the 
narrowest part (the weakest bond) of the channel. With some reasonable 
assumptions, one \cite{Halperin et al:1985_SCR} can express the transport
capacities of the weakest region in terms of percolation cluster statistics 
on the lattice (particularly in terms of the percolation correlation 
length\index{correlation length} $\xi_p$).

Both the discrete and continuum type are almost same in the dilute 
limit ($p \to 1$). 
So all the results derived earlier for discrete model are valid here.
 But near criticality an infinite percolation 
 cluster, with the links of mean length $\xi_p$
and of different cross-sectional width $\delta$, is formed.
The backbone \cite{de Gennes:1976_SCR,Skal and Shklovskii:1975_SCR} of this 
cluster is represented by a superlattice (see sec.~\ref{sec:criticality_SCR}) 
of 
tortuous singly connected links and blobs crossing at nodes at a separation 
\begin{equation}
\label{eqn:si_SCR}
\xi_p \sim |p-p_c|^{-\nu}.
\end{equation}
The chemical length\index{chemical length} between any two nodes is 
\begin{equation}
\label{eqn:lc_SCR}
L_c \sim |p-p_c|^{-\zeta}. 
\end{equation}
For singly connected bonds (or sites) on the percolating backbone, $\zeta=1$  
in all dimensions\index{exponent!chemical length} 
\cite{Halperin et al:1985_SCR} except $d=1$ 
\cite{Stauffer and Aharony:1992_SCR, 
Coniglio:1982_SCR}. There are $(L/\xi_p)^{d-1}$ number of parallal links exist 
between two electrodes,
where $L$ is the size of the sample. The current in a link is given 
by $i_L=(\xi_p/L)^{d-1}I$. The current density $i$ in a channel of 
cross-section $\delta$ is given by 
\begin{equation}
i \sim \frac{\xi_p^{d-1} I}{\delta^{d-1}}
\end{equation}   
The maximum current density is obtained for minimum 
cross-section $\delta_{min}$ which is inversely proportional to the shortest 
chemical 
length $L_c$ \cite{Halperin et al:1985_SCR,Baudet and:1985_SCR,Benguigui:1986_SCR}.
 If $i_0$ is the threshold current density at which the material fails, then

\begin{equation}
i_0 \sim \xi_p^{d-1} L_c^{d-1}I_f
\end{equation} 
and we get,

\begin{eqnarray}
I_f \sim (p-p_c)^{\nu+1} \hspace{5mm} (\mbox{in} \hspace{1mm} 2d),\hspace{10mm}I_f \sim (p-p_c)^{2(\nu +1)} \hspace{5mm} (\mbox{in} \hspace{1mm} 3d).
\end{eqnarray}

Thus the exponents\index{exponent!fuse failure current in continuum model} 
for the failure current are higher than those of
the discrete model.

Using $I_f=V_f/R$ and $R \sim |p-p_c|^{-\tilde{t_c}}$ with
equations~(\ref{eqn:si_SCR}) and (\ref{eqn:lc_SCR}) the expression for fuse 
voltage becomes $V_f \sim |p-p_c|^{\tilde{t}_f}$ with 
$\tilde{t}_f$ equal to $(d-1)(\nu + 1)-\tilde{t_c}$ \cite{Chakrabarti:1997_SCR}.
$\tilde{t}_c$ is the 
conductivity\index{exponent!conductivity in continuum model} 
exponent: $\tilde{t}_c \simeq1.3$ 
and $2.5$ in $2d$ and $3d$ respectively. Consequently, $\tilde{t}_f$ 
becomes approximately\index{exponent!fuse failure voltage in continuum model} 
$1$ and $1.3$ in $2d$ and $3d$ respectively.
Unlike discrete percolation where failure current at $p_c$ attains a finite 
value in $2d$ and vanishes only in $3d$, in continuum percolation failure 
voltage always vanishes at $p_c$.   

\begin{table}
\caption{\footnotesize{Theoretical estimates\index{exponent!fuse 
          failure current in discrete model} for the fuse 
          failure exponent\index{exponent!fuse failure current in 
          continuum model} $\phi_f$}}
\label{tab:fuseexponent_SCR}
\vspace{10pt}
\centering
\begin{tabular}{|c|c|c|}
\hline
Dimension&Lattice percolation&Continuum percolation\\
\hline
$2$&$\nu(=4/3)$&$\nu+1(=7/3)$\\
\hline
$3$&$2\nu(\simeq 1.76)$&$2(\nu+1)(\simeq 3.76)$\\
\hline
\end{tabular}
\end{table}


%% file: mig.tex
\label{sec:shortest_SCR}
Electromigration is an example where we find the practical application of the 
concepts we have hitherto discussed.
Miniaturization of the circuits and gadgets have become the norm of the day. 
Very thin metallic films are used as interconnects among the 
active parts of devices, resulting in large current densities through
 the metallic films. Having large enough momentum, the free electrons 
become able to displace metallic ions from their equillibrium positions.
Thus depending upon the material a net material transport 
occours \cite{K. N. Tu:2003_SCR} through grain boundary diffusion, surface 
diffusion or lattice diffusion. This ionic displacement and 
the accumulated effect of material transport due to large 
current densities (more than $10^4$ $A/cm^2$) are 
called \textit{electromigration}\index{electromigration}.   
	Over time such electromigration leads to void formation at 
the cathode and extrusion at the anode in thin film interconnects.
 Such void formation and hillock formation cause an open circuit 
and a short circuit to neighbouring connecting wires respectively. 
The problem of material accumulation can be suppressed by 
the layers of other material around and above the interconnects. 
The problem of open circuit due to void formation has received 
much attention as the random failures of several interconnects 
finally lead to failure of the entire device.

Let a fraction $(1-p)$ of the resistors are removed 
from random resistor network (say in 2d). A random walker 
starts its walk from one of the lateral sides (those without electrodes) 
of the network, and jumps from one cell to a nearest neighbour 
cell by crossing the bonds irrespective of occupied or 
unoccupied bonds, intending to reach the opposite side, with the 
constraint that it never visits a cell more than once.  		 
Let the whole path consists of $n_0$ occupied and $n_1$ unoccupied 
bonds. For a given configuration of missing resistors the minimum 
value of $n_0$ is the \textit{shortest path}\index{shortest path}.The mean 
shortest path $<n_0>$ (for theoretical study see 
ref.~\cite{Chayes et al:1986_SCR,Stinchcombe et al:1986_SCR}) can be realised 
by 
considering the average over a large number of configurations.
An electromigration induced failure of the network can be realised 
if the walker fuses the resistors as it crosses them.
 The criterion for a resistor to fuse is not a particular value of 
current (as in the fuse model), rather a particular value of thershold 
charge $Q_0$ that attains the resistor from the time of 
application of the constant current $I_0$ to the network (for details 
of the model 
see ref.~\cite{Bradley and Wu:1994_SCR,Wu and Bradley:1994_SCR}). If $t_1$ is 
the failure time of an arbitrary resistor of the network 
which is subjected to a constant current $I_0$, then the 
resistor must satisfy
\begin{equation}
\int^{t_1}_0 I(t)dt=Q_0
\end{equation}
to fail.

 So \textit{shortest path}\index{shortest path} for a given impurity 
concentration is the path 
which corresponds to 
the smallest number of resistors to fuse. 
The problem is to study the variation of failure time of the 
whole network $\tau$ with $p$.
For a network having an isolated defect of length $n$, $\tau$ is given by
\begin{equation}
\tau=(L-n)Q_0/I_0,
\end{equation}
where $LQ_0/I_0$ is the $\tau$ value for the pure network. 
In pure limit the mean failure time $<\tau>$ is related to the 
longest probable defect with size $n_c$. This means that the bonds 
which fuse are also the bonds with largest current.

$\tau$ decreases as $p$ decreases from pure limit, and approaches zero 
at $p_c$, since the number of bonds to be broken goes also to zero. 
The shortest path $n_0$, the failure current $I_f$, the number of broken bonds
$N_f$ and the failure time $<\tau>$, all these quantities tend to go to zero 
at $p_c$ with the correlation length 
exponent\index{exponent!correlation length} $\nu$. 


%% file: fusen.tex

The numerical simulation of failure of random fuse network  
in $2d$ was carried out by de 
Arcangelis \textit{et al.} \cite{de Arcangelis et al:1985_SCR}.
Here, one starts with a lattice, the bonds of which are conductors and 
present with a probability $p$. An increasing external voltage is applied 
across the lattice and the voltage $V_f^1$ when there is first local failure is 
recorded. The fused bond is removed and the voltage is raised till there 
is a second failure  at $V_f^2$. The process is continued till the global 
failure at a voltage $V_f^{fin}$ occurs. The variation of $V_f^1$ 
and $V_f^{fin}$ with $p$ are studied. $V_f^1$ decreases with decreasing $p$ 
till $p \sim 0.7$ whereupon it attains minimum value and then 
it starts increasing again. $V_f^{fin}$, on the contrary, increases
monotonically and behaves almost identically as $V_f^1$ for $|p-p_c|<0.08.$ 
Both approaches $p_c$ with the diverging 
exponent $0.48$\index{exponent!fuse failure voltage in discrete model}. 
It seems that $V_f^1$ exhibits pseudodivergence in $2d$.
  
Duxbury \textit{et al.} \cite{Duxbury et al:1987_SCR} performed 
similar type of simulations 
covering the whole range of $p$ from $1$ to $p_c=0.5$. Their 
simulation results concerning $I_f$ vs. $p$ graph fit well with 
the iterpolation formula~(\ref{eqn:unifiedif_SCR}) with the exponent 
value $\phi_f=1$\index{exponent!fuse failure current in discrete model}, 
whereas the theoretically predicted value is $1.33$. 
This may be due to the finite size effect.

\begin{figure}[htbp]
   \begin{center}
      \includegraphics[width=0.60\textwidth]{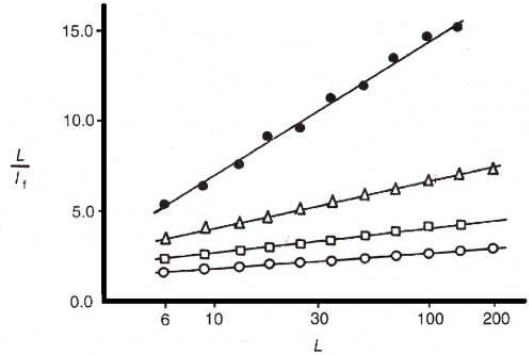}\hspace{2mm}
      \caption{\footnotesize{$L/I_f$ vs. $\ln L$ expressing their 
              linear dependence (after 
              Duxbury \textit{et al.} \cite{Duxbury et al:1987_SCR}). 
              The curves from top to 
              bottom correspond
              to initial impurity probability $p=0.6,0.7,0.8$ 
              and $0.9$ respectively.}}
      \label{fig:ifl_SCR}
   \end{center}
\end{figure}

They also demonstrated the finite value 
of $V_f^1$\index{exponent!fuse failure voltage in discrete model} at $p_c$ by looking into 
the variation of $I_f$ and the conductance near $p_c$.
Following equation~(\ref{eqn:ifp1_SCR}) they checked the linear dependence 
of $L/I_f$ on $\ln L$ successfully for several values of $p$ varying $L$ 
from $10$ to $200$ and 
determined $\psi_f=1$\index{exponent!size dependence of fuse failure current} 
from their slopes. The slope 
increases as $p$ approaches $p_c$.
Instead of~(\ref{eqn:fibk1_SCR}) they preferred
\begin{equation}
\label{eqn:fidux_SCR}
F_L (V)=1-exp\left [ -AL^2 exp\left ( -\frac{KL}{V} \right ) \right ]
\end{equation}
as\index{distribution function!Gumbel} cumulative distribution function of failure voltage.

Arcangelis \textit{et al.} \cite{de Arcangelis et al:1985_SCR} and 
Duxbury \textit{et al.} \cite{Duxbury et al:1987_SCR} 
determined the number of fused bonds $N_f$ upto complete failure and they 
found that $N_f$ 
goes\index{exponent!number of broken bonds upto global fuse failure} to zero algebraically as $p \to p_c$ with an exponent 
seemingly equal to the correlation length exponent $\nu$.

%% file: di.tex
In dielectric breakdown\index{dielectric breakdown} 
problem the sample is insulating and
conducting material acts as defects in the composite. 
The volume fraction ($p$) of the conducting material is less than the 
critical volume fraction ($p_c$) so that the system overall is not conducting. 
The dielectric portions can withstand electric field upto $e_c$, 
at and beyond which they become conducting (local breakdown). 
Global breakdown occurs when the conducting portions
 span the sample with the succession of local breakdown
under the influence of an increasing external field. 

%

In $2d$ the solution for dielectric problem can be  obtained from the solution 
of fuse problem with the aid of \textit{duality relation}
 \cite{Mendelson:1975_SCR, Bowman and Stroud:1989_SCR}.
We follow Bowman and Stroud \cite{Bowman and Stroud:1989_SCR} and 
consider the case where $p$ is smaller 
than $p_c$ and the sample is macroscopically insulating.  
 The equations for the induction vector $\mathbf{D}$ and the field $\mathbf{E}$ are:
\begin{eqnarray}
\label{eqn:max1_SCR}
\bigtriangledown \cdot \mathbf{D}  =  0 \nonumber \\
\bigtriangledown \times \mathbf{E}  =  0 \\
\mathbf{D}(r)  =  \epsilon (r) \mathbf{E}(r), \nonumber
\end{eqnarray}
where $\epsilon (r)$ is local dielectric constant of the insulating portion and
$\mathbf{E}$ is irrotational .
With $\mathbf{E}=- \bigtriangledown \phi$ equations~(\ref{eqn:max1_SCR}) yields
    
\begin{equation}
\label{eqn:phi_SCR}
\frac{\partial}{\partial x} \left \lbrack \epsilon (r) \frac{\partial \phi}{\partial x}\right \rbrack
+ \frac{\partial}{\partial y} \left \lbrack \epsilon (r) \frac{\partial \phi}{\partial y}\right \rbrack
=0,
\end{equation}
where $\phi$ is the scalar potential.
Now consider the dual composite of the original, where insulator 
phase of the original is replaced by conducting phase and vice versa. The 
sample is macroscopically conducting and the relevant quantities are 
current density $\mathbf{i}$ and electric field $\mathbf{\bar{E} }$.
These satisfy the equations
\begin{eqnarray}
\label{eqn:max2_SCR}
\bigtriangledown \cdot \mathbf{i}  =  0 \nonumber \\
\bigtriangledown \times \mathbf{\bar{E}}  =  0 \\
\mathbf{i}(r)  =  \sigma (r) \mathbf{\bar{E}}(r) \nonumber
\end{eqnarray}
Since $\mathbf{i}$ is divergenceless, $\mathbf{i}$ can be expressed as a curl 
of a vector potential $\mathbf{V}$.
$\mathbf{V}$ ($\mathbf{V}=V_{z}=\psi(x,y)$) is chosen in such a way, 
only z-component of $\mathbf{i}$ vanishes.   
 Assuming local conductivity as $\sigma (r) = 1 / \epsilon (r)$ 
equations~(\ref{eqn:max2_SCR}) yields
\begin{equation}
\label{eqn:psi_SCR}
\frac{\partial}{\partial x} \left \lbrack \epsilon (r) \frac{\partial \psi}{\partial x}\right \rbrack
+ \frac{\partial}{\partial y} \left \lbrack \epsilon (r) \frac{\partial \psi}{\partial y}\right \rbrack
=0.
\end{equation}
Comparing (\ref{eqn:phi_SCR}) and (\ref{eqn:psi_SCR}), one has 

\begin{equation}
\frac{\partial \psi}{\partial x} = \frac{\partial \phi}{\partial x},
\hspace{5mm}
\frac{\partial \psi}{\partial y} = \frac{\partial \phi}{\partial y}.
\end{equation}
With all these, the components of current density $\mathbf{i}$ 
in dual composite become
\begin{equation}
i_x=\partial \psi / \partial y = E_y,
\hspace{5mm}
i_y=-\partial \psi / \partial x = -E_x.
\end{equation}

Thus we see that the magnitude of the current density in the dual composite 
is equal to the magnitude of the electric field in the
 original one and the direction is rotated by $90^0$.
It can also be shown that the field $\mathbf{\bar{E}}$ of the dual problem
is equal to vector $\mathbf{D}$ of the dielectric problem and is rotated by 
$90^0$. Now the physical correspondence of the two pictures
 can easily be established \cite{Chakrabarti:1997_SCR,Sahimi:2002_SCR}.

All the results of fuse problem can be utilized for obtaining 
the solution for dielectric problem. For example, 
the equations (\ref{eqn:unifiedif_SCR})
and (\ref{eqn:fipc_SCR}) can be used in a straight forward way on replacing 
$I_0$ and $I_f$ by $V_0$ and $V_b$ respectively and also $(1-p)$ by $p$.


%% file: tdi.tex
\subsection{Dielectric breakdown problem}

As in the fuse problem, dielectric breakdown\index{dielectric breakdown} 
problem can be viewed
on a discrete lattice with disorder in the framework of percolation theory.
Basically, a lattice of insulating bonds is considered out of which
$p$ $(p<p_c)$ fraction of bonds are conducting at random. For this 
purpose resistor (capacitor) of different resistivity (capacitance)
\cite{Bowman and Stroud:1985_SCR, Bowman and Stroud:1989_SCR} (\cite{Beale and Duxbury:1988_SCR}) 
can be used. Resistors (capacitors) of smaller (higher) resistance (capacitance)
are used for conducting bonds. 

\subsubsection{Dilute limit ($p \to 0$)}
 
The problem is very similar to the fuse problem.
The enhanced local field due to the presence of long 
defect (made of $n$ number of larger valued 
capacitors (conducting bonds) perpendicular
to the electrode, 
is (following  Beale and Duxbury \cite{Beale and Duxbury:1988_SCR})
\begin{equation}
E_e = E(1+kn),
\end{equation}
where $E$ is the externally applied field and $k$ is the enhancement factor.
It is valid for all dimension
The probability to find a long defect made of $n$ conducting bonds is

\begin{equation}
P(n) \sim p^nL^d.
\end{equation}
The most probable defect size is 
\begin{equation}
n_c=-\frac{d}{\ln p}\ln L.
\end{equation}
The enhanced field near the insulating bond adjacent to 
the most probable defect is now 
\begin{equation}
E_e =E \left \lbrack 1+K_d \left (-\frac{\ln L}{\ln p} \right ) \right \rbrack. 
\end{equation}
When enhanced field $E_e$ of any bond attains threshold 
value of dielectric breakdown, 
local breakdown takes place and the bond becomes conducting.
So breakdown voltage becomes
\begin{equation}
V_b = \frac{E_0 L}{ \left \lbrack 1+ K_d \left ( - \frac{\ln L}{\ln p}\right ) \right \rbrack},
\end{equation}
where $E_0$ is the breakdown field without defect.

\subsubsection{Close to critical point ($p \to p_c$)}
\label{sec:critical_SCR}
In this limit, consider the 
conducting defects which are on the average the percolation correlation length
$\xi_p$ distance apart. Hence, average field is $V_1/\xi_p=V/L$, where 
$V$ and $V_1$ are externally applied voltage and potential difference 
between any two conducting defects respectively. The maximum attainable 
field is $V_1/a$, where $a$ 
is the minimum available separation (bond length) between the defects 
in the network. When the local
electric field $V_1/a$ reaches the bond-threshold value $e_c$, breakdown occurs.
Now the required average electric field for local breakdown is $E_b=(a/\xi_p)e_c$.
Near $p_c$, $\xi_p$ diverges\index{exponent!correlation length} 
as $\xi_p(p) \propto a(p-p_c)^{-\nu}$.  
So the critical behavior of the average breakdown 
field as derived by  Lobb \textit{et al.} \cite{Lobb et al:1987_SCR} 
appears as\index{exponent!dielectric breakdown field in discrete model}
\begin{equation}
E_b \sim (p_c - p)^{\nu}.
\end{equation}
in all dimensions. This can also be 
derived \cite{Stinchcombe et al:1986_SCR} from the concept of 
minimumgap\index{minimum gap} $g(p)$, as $E_b$ can be considered 
as proportional (see~sec.~\ref{sec:shortest_SCR}) to $g(p)$ and $g(p)$ is 
proportional to $\xi_p^{-1}$\index{exponent!minimum gap} 
(see relation~\ref{eqn:si_SCR}), the result above 
follows. Since $E_b$ is inversely proportional to the linear dimension of the 
most vulnerable defect and the typical size of the defect or conducting 
clusters diverges at $p_c$, $E_b$ approaches zero in this limit.
                                                                                                                                                                                       
\subsubsection{Influence of sample size}

We consider a critical defect which is a  pair of 
longitudinal very closely spaced clusters of  conducting 
material (almost linear in shape) of the order of size $l$ situated one after 
one, parallal to the direction of the applied field $E$
in a $L^d$ lattice. The field between these two clusters enhanced by 
a factor $l$ time $E$. Far from $p_c$ these clusters appear 
with the probability $(1/\xi_p)exp(-l/\xi_p)$. Thus the most dangerous 
defect (the largest cluster) is of the order of 
$l_{max} \sim \xi_p \ln (L^d)$. 
The first breakdown field $E^1_b$ (field needed to break the first bond) 
scales as $1/l_{max}$ in a finite but large 
system \cite{Beale and Duxbury:1988_SCR}. Since, in this particular case breakdown 
field for the sample is same as for the first breakdown, $E_b$ 
becomes\index{exponent!dielectric breakdown field in discrete model}
\begin{equation}
\label{eqn:ebsize_SCR}
E_b \sim \frac{(p_c -p)^{\nu}}{\ln L}.
\end{equation}   
This approximate expression for the average breakdown field was derived  
by Beale and Duxbury \cite{Beale and Duxbury:1988_SCR}.

Larger sample has higher chance of having a larger vulnerable
defect. So larger sample requires smaller electric field for breakdown 
and $E_b$ becomes zero in the limit $L\to\infty$. Breakdown path is always on 
the average perpendicular to the electrode. So the above result is independent 
of dimension.

Due to the dominating behavior of $(p_c-p)^\nu$ over $\ln L$, 
breakdown field tends to zero as $p$ approaches $p_c$.
Very near to $p_c$ Chakrabarti and Benguigui (\cite{Chakrabarti:1997_SCR} p. 66) 
proposed a scaling relation as  
\begin{equation}
E_b \sim\frac{(p_c-p)^\nu}{\ln (L/\xi_p)}.\index{exponent!dielectric breakdown field in discrete model}
\end{equation}
According to Bergman and Stroud \cite{Bergman and Stroud:1992_SCR}, $E_b$ may become size independent
very close to $p_c$, and a cross-over from extreme statistics to percolation 
dominated statistics may be there.

\subsubsection{Summary}
\label{sec:unify_SCR}
A general formula may be inferred for dielectric breakdown field as:
\begin{equation}
\label{eqn:interpolation_SCR}
E_b=E_0\frac{[(p_c-p)/p_c]^{\phi_b}}{1+K \left \lbrack \frac{\ln (L/\xi_p)}{-\ln p}\right \rbrack},
\end{equation}  
with the exponent $\phi_b$ (=$\nu$ for lattice 
percolation)\index{exponent!dielectric breakdown field in discrete model} 
dependent on the 
dimension and on the type of percolation.
For the theoretically estimated value for $\phi_b$ 
see table~\ref{tab:dielectricexponent_SCR}.

\begin{table}
\caption{\footnotesize{Theoretical estimates for the dielectric 
         breakdown exponent\index{exponent!dielectric breakdown field in 
         discrete model} $\phi_b$}}
\label{tab:dielectricexponent_SCR}
\vspace{10pt}
\centering
\begin{tabular}{|c|c|c|}
\hline
Dimension&Lattice percolation&Continuum percolation\\
\hline
$2$&$\nu(=4/3)$&$\nu+1(=7/3)$\\
\hline
$3$&$\nu(\simeq 0.88)$&$\nu+1(\simeq 1.88)$\\
\hline
\end{tabular}
\end{table}

\subsection{Distribution of breakdown field}

Distribution function $F_L(E)$ in dielectric breakdown is defined as the 
probability of breakdown of a dielectric system of size $L$
in an external electric field $E$.
Usually, Weibull distribution\index{distribution function!Weibull} 
\cite{Weiderhorn:1984_SCR,Weibull:1961_SCR}
\begin{equation}
\label{eqn:WeibullE_SCR}
F_L(E)=1-exp(-rL^dE^m)
\end{equation}
is used to fit the distribution function of breakdown and failure 
problems. Here $r$ and $m$ are constants. Duxbury and his co-workers
\cite{Duxbury et al:1986_SCR,Duxbury et al:1987_SCR,Duxbury and Leath:1987_SCR,Beale and Duxbury:1988_SCR} 
argued that the distribution function for dielectric (and electrical) breakdown
is given by Gumbell distribution\index{distribution function!Gumbel}:
\begin{equation}
\label{eqn:BDE_SCR}
F_L(E)=1-exp \left \lbrack -AL^d exp \left ( - \frac{K}{E} \right ) \right \rbrack.
\end{equation} 
It can be derived \cite{Beale and Duxbury:1988_SCR} from a simple scaling arguement
based on percolation cluster statistics \cite{Stauffer:1979_SCR,Essam:1980_SCR,Kunz and Souillard:1978_SCR}. 
It is valid for $L>>\xi_p$ or in dilute limit.
The two expression (\ref{eqn:WeibullE_SCR}) and (\ref{eqn:BDE_SCR}) 
are qualitatively 
same if the Weibull exponent $m$ is large. Though second one provides
better fit to numerical simulation \cite{Beale and Duxbury:1988_SCR}. 
Sornette \cite{Sornette:1988_SCR} argued that 
the expression~(\ref{eqn:BDE_SCR}) 
is not appropriate 
in continuum system having percolation type of disorder, rather, Weibull-like 
distribution is valid there.

\subsection{Continuum model}
The model consists of dielectric material with the spherical (circular in $2d$) 
conducting inclusions as defects with randomly positioned centres 
having the possibility to overlap. We assume breakdown
field $E_b$ is proportional to the width $\delta$. For breakdown $\xi_p^{-1}$ 
numbers of link elements to be broken. With the same logic
as in sec.~\ref{sec:continuum_SCR}, $E_b \sim \delta_{min} \xi_p^{-1} \sim L_c^{-1} \xi_p^{-1}$ or  
\begin{equation}
E_b\sim (p_c - p)^{\nu+1}\index{exponent!dielectric breakdown field 
in continuum model}
\end{equation} 
very near to $p_c$ \cite{Chakrabarti et al:1988_SCR,Lobb et al:1987_SCR}. So 
continuum system is weaker than the 
discrete one. This is because the conductivities of the  conducting channels 
increases as $p\to p_c$, whereas, conductivity of the bonds
in discrete model is independent of $p$.

\subsection{Shortest path}
The concept of shortest path\index{shortest path} is already mentioned 
in section~\ref{sec:shortest_SCR} in the context of electromigration.
Just like before, we consider a walker starting its 
walk from one electrode with jump from one 
site to another and reaches the opposite electrode after 
executing a self-avoiding walk. The aim of the walker is to create 
a percolating conducting path by transforming a 
insulating bond to conducting bond when it jumps between two conducting sites 
separated by the insulating bond. $n_0$ denotes the total 
number of insulating bonds 
which are to be transformed into conducting bonds during walk across the 
sample. After completion of the walk, the sample experiences a
continuous conducting path between the electrodes and in this way
dielectric breakdown  comes into existance.

The shortest path\index{shortest path} for a given configuration is defined as 
the path with the smallest $n_0$.The normalised smallest path 
is given by $g(p)=<n_0>/L$, where $<n_0>$ is obtained by
considering the average of $n_0$ over a large number of configurations. 
Some authors call $g(p)$ by minimum gap\index{minimum gap}.

The behavior of $g(p)$ has been studied \cite{Stinchcombe et al:1986_SCR,
Duxbury and Leath:1987_SCR} extensively in $2d$ and $3d$ for 
regular and directed 
percolation scenario. As $p$ increases from zero to $p_c$, the minimum 
gap $g(p)$ decreases from unity to zero with the correlation length 
exponent\index{exponent!correlation length} $\nu$.


%% file: sto.tex
\subsection{Numerical simulations in dielectric breakdown}

\bigskip

\textbf{Stochastic models}

\bigskip

In these models, stochastic growth processes are considered which
mimic the dielectric breakdown processes. For example, 
Sawada \textit{et al.} \cite{Sawada et al:1982_SCR} considered 
a random growth process where 
growth takes place in two ways: tips  of the pattern grow with a probability 
$p_0$ and new tips (branching) appear with probability $p_n$ (here, $p_n<p_0)$. 
In their simulation the pattern appears as fractal and fractal dimension can 
be tuned by the parameter $R$ $(=p_o/p_n)$. However, results of such a 
simplification do not satisfy experimental results of dielectric 
breakdown \cite{Niemeyer and Pinnekamp:1982_SCR}.

To mimic the dielectric discharge pattern in gasses 
Niemeyer \textit{et al}. \cite{Niemeyer et al:1984_SCR} suggested 
a stochastic model where the 
breakdown pattern generated in turn determines the local electric field and 
the growth probability. The model could reproduce the fractal properties of 
dielectric breakdown process by numerical simulations.
In this model the breakdown patttern starts growing from the centre of a 
lattice with insulating bonds. One electrode is placed at the centre and 
other one is placed at a long distance on the circumference of a circle. In 
one step only one interface bond (and a point) among all the nearest neighbors 
of the pattern breaks down depending upon the growth probability $p$ and 
becomes the member of the pattern. The newly added bond becomes a conductor 
and the newly added site is shorted to the voltage of the central electrode. 
The growth starts from the centre and grows radially outwards. The growth 
probability $p$ depends on the local field (potential) which in turn is 
controlled by the breakdown pattern via the relation 

\begin {equation}
p(i,j \to i \prime , j \prime ) = \frac{(V_{i \prime , j \prime})^{\eta}}{\sum (V_{i \prime , j \prime})^{\eta}},
\end {equation} 
where the indices $i,j$ and $i \prime, j \prime$ represent the discrete 
lattice coordinates.
The electric potential $V$ is defined for all sites of the lattice by the 
discrete Laplace equation\index{Laplace equation} 

\begin{equation}
\label{eqn:laplace_SCR}
V_{i,j} = \frac{1}{4} (V_{i+1,j} + V_{i-1,j} + V_{i,j+1} + V_{i,j-1})
\end{equation}      
with the boundary condition $V=0$ for each point of the discharge pattern 
and $V=1$ outside the external circle. $\eta$ describes the relation between 
local field and growth probability. The fractal structure of the pattern 
has been seen to obey

\begin{equation}
N(r) \sim r^{d_f},
\end{equation}  
where $N(r)$ is the total number of discharge points inside a circle of 
radius $r$ and $d_f$ is the Hausdorff dimensions\index{Hausdorff dimension}. 
In $2d$ one 
has $d_f \simeq 2,1.89 \pm 0.01, 1.75 \pm 0.02, 1.6$ 
for $\eta = 0, 0.05 ,1, 2$ respectively. The structure tends to be more linear
with larger $\eta$. The observed value of $d$ for $\eta =1$ is in
 good agreement with the experimental result $(\simeq 1.7 )$ (Niemeyer 
and Pinnekamp \cite{Niemeyer and Pinnekamp:1982_SCR}) and the resulting 
figure is very similar to Lichtenberg 
figure.

The same pattern for $\eta =1$ has been produced surprisingly by a different
 growth model: diffusion-limited aggregation model (DLA) of 
Witten and Sadner \cite{Witten and Sander:1981_SCR} 
(For review see Meakin \cite{Meakin:1998_SCR}).    
Niemeyer \textit{et al.} neither justifies the appearance of $\eta$ nor 
provides any theoretical explanation regarding explicit rule for breakdown.
Many models (for details see ref.~\cite{Sahimi:2002_SCR}) have been 
proposed but a model which fully describes dielectric breakdown in solids is 
still lacking. 

\bigskip

\noindent \textbf{Deterministic models}

\bigskip

Contrary to Niemeyer \textit{et al}. \cite{Niemeyer et al:1984_SCR}, 
Takayasu \cite{Takayasu:1985_SCR} introduced 
deterministic approach to produce dendritic fractal pattern (as is found 
in lightning) by considering \textit{a priori} spatial fluctuation on
bond resistances ($r_i=\theta r;\hspace{1mm} \theta \in [0,1]$) and 
the nonlinear 
irreversible characteristics of the resistance: once the potential difference 
across a resistance $r_i$ attain a pre-assigned 
threshold voltage\index{threshold!dielectric breakdown voltage} 
$v_c$, $r_i$ reduces 
to $\delta r_i$ ($\delta$ is a small positive quantity) and 
it's resistivity then never changes. This is the breakdown of a resistor in 
this model. Breakdown of a resistor induces successive breakdown of other 
resistors leading to the formation of percolation cluster of broken resistors. 
The pattern appears to be anisotropically fractal with the dimension 
$d_f=1.58 \pm 0.12$ in $2d$. Due to anisotropy this $d_f$ is less than the 
fractal dimension 1.89 of a percolating network in $2d$.

Family \textit{et al.} \cite{Family et al:1986_SCR} made 
the stochastic model of  
Niemeyer \textit{et al.} \cite{Niemeyer et al:1984_SCR} deterministic 
by attaching randomly a breakdown 
coefficient $\theta$ ($\theta\in[0,1]$) to each insulating 
bond. There are two versions of this model: In one model, at each time step
an interface bond $ij$ with the largest $\theta V_{ij}^\eta$ breaks down,
whereas in the other model an interface bond breaks down with a probability
$\theta V_{ij}^{\eta} /p_{max}$, where $\eta$ is an adjustable constant 
and $p_{max}$ is the largest value of $\theta V_{ij}^{\eta}$ among all
the interface bonds. The patterns appear as tunable (varying $\eta$) fractal.
In the first model stringy highly anisotropic (with branches) pattern
with fractal dimension of about $1.2$ appears in $2d$, whereas in 
the second one pattern is strikingly lacking in anisotropy. The shape
of the pattern is similar to those found in  
Niemeyer \textit{et al.} \cite{Niemeyer et al:1984_SCR} 
except the fractal dimension of $1.70\pm0.05$ (for $\eta=1$) in $2d$. The 
results for $d_f$ appear same to that for DLA on a square lattice
\cite{Meakin:1985_SCR} and to that for the same model in a homogeneous 
medium (here, cross and square-shaped patterns appear)
 \cite{Family et al:1986_SCR}.  
 
Direct simulations of dielectric breakdown from various groups confirmed the  
theoretical consideration for size and impurity dependence of breakdown voltages
quite successfully.  In these simulations, a discrete lattice
of insulating bonds is considered with $p$ fraction of bonds as conductors. 
Each insulator breaks down to conductor at a threshold value of voltage drop
$v_c$. A macroscopic voltage is applied 
and the voltage distribution throughout the lattice is computed by solving 
Laplace equation~(\ref{eqn:laplace_SCR})\index{Laplace equation}. 
The insulator with largest voltage drop
(at or above $v_c$) is converted to a conductor (incidence of first local 
breakdown). The voltage distribution is then recalculated, the second local 
breakdown is identified and the process continues. If at any step the applied 
voltage is not large enough to cause breakdown of any insulator, it is 
increased gradually. The simulation continues till a sample spanning cluster of
conductors appears. The applied voltage $V_b$ at which global breakdown occurs 
divided by the sample size $L$ is identified as breakdown field $E_b$.
It has been experienced that the breakdown field $E_b$ is same as the field
$E_b^1$ required for the occurrence of first local breakdown.

\begin{figure}[htbp]
   \begin{center}
      \includegraphics[width=0.60\textwidth]{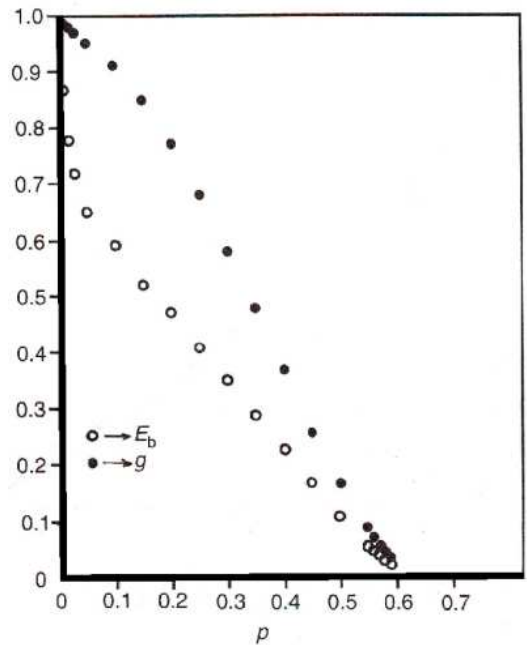}\hspace{2mm}
      \caption{\footnotesize{The variation of first breakdown 
              field\index{dielectric breakdown!field} $E_b$ and the 
              minimum gap\index{minimum gap} 
              $g(p)$ with initial impurity 
              probability $p$, (after Manna and Chakrabarti 
              \cite{Manna and Chakrabarti:1987_SCR}). The 
              two quantities behave same very near to the percolation 
              threshold $p_c$, whereas far from $p_c$ they behave differently.}}
      \label{fig:ebgp_SCR}
   \end{center}
\end{figure}

Manna and Chakrabarti \cite{Manna and Chakrabarti:1987_SCR} determined 
the $p$ dependence of both $E_b^1$ 
and $g(p)$ for the entire range of $p$ below $p_c$. They found 
(see figure~\ref{fig:ebgp_SCR}) that both $E_b^1$\index{exponent!dielectric 
breakdown field in discrete model} and 
$g(p)$\index{minimum gap} go to zero at
$p_c$ with the exponent of value almost equal to 1. They argued that 
the exponent is actually $\nu$ and the smallness of the value is due 
to the smallness of the lattice size ($L=25$, where exact $p_c$ is not 
reachable). Bowman and Stroud \cite{Bowman and Stroud:1989_SCR} found 
that $E_b^1$ vanishes at $p_c$ with 
an exponent\index{exponent!dielectric breakdown field in discrete model} 
equal to $1.1 \pm 0.2$ in $2d$ and $0.7 \pm 0.2$ in $3d$ 
for both site and bond problems. These are consistent with the correlation 
length exponent\index{exponent!correlation length} ($\nu$) values of $4/3$ 
and $0.88$ in $2d$ and $3d$ respectivey.				
 
Beale and Duxbury \cite{Beale and Duxbury:1988_SCR} proposed a 
relation for first (local) 
breakdown field as
\begin{equation}
E_b^1 \sim \frac{1}{A(p)+ B(p)\ln L}.
\end{equation}
  Near $p_c$, they found $A(p)$ and $B(p)$ vary like $(p_c-p)^{-\nu}$ 
(in the lattice of size $L=50,70,100$) as is expected from
eqn.~(\ref{eqn:ebsize_SCR}). From the plot 
of  $\ln \lbrace B(p)\ln (p) \rbrace$ 
versus $-\ln (p_c-p)$ their data shows $\nu = 1.46 \pm 0.22$ which is 
in good agreement with exact value $4/3$ in $2d$. Their data fits well to the 
Gumbel\index{distribution function!Gumbel} double exponential 
form (\ref{eqn:BDE_SCR}) for cumulative failure 
distribution.  
Manna and Chakrabarti \cite{Manna and Chakrabarti:1987_SCR} and 
Beale and Duxbury \cite{Beale and Duxbury:1988_SCR} found the 
exponent $\phi_b$\index{exponent!dielectric breakdown field in discrete model} 
of eqn.~(\ref{eqn:interpolation_SCR}) to be 
about $1.0$ and $1.2$ respectively.

Acharyya and Chakrabarti \cite{Acharyya and Chakrabarti b:1996_SCR} found 
that starting with a concentration $p$ $(p<p_c)$ 
of the conductors, the rate at which the insulating bonds breaks down to 
conductors as the electric field is raised in a dielectric breakdown diverges 
at breakdown voltage $V_b$.

\begin{figure}[htbp]
   \begin{center}
      \includegraphics[width=0.60\textwidth]{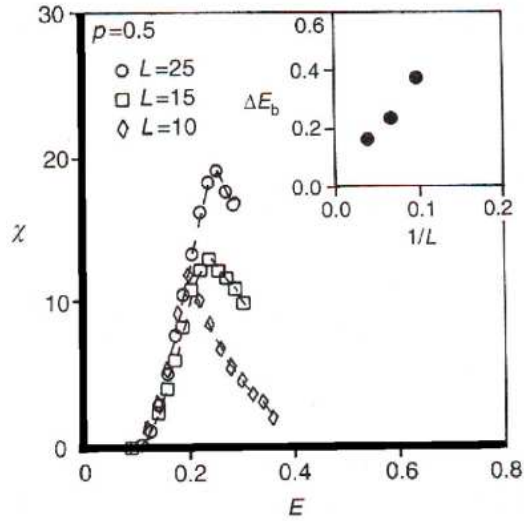}\hspace{2mm}
      \caption{\footnotesize{The plot of dielectric breakdown 
              susceptibility\index{breakdown susceptibility} vs. applied 
              voltage for several sample sizes (after 
              Acharyya and 
              Chakrabarti \cite{Acharyya and Chakrabarti a:1996_SCR}).
              The inset shows that the difference 
              of the maximum susceptibility field 
              and the minimum sample spanning cluster field decreases with 
              increasing sample size}}
      \label{fig:susceptiblity_SCR}
   \end{center}
\end{figure}
 This indicates that 
 global breakdown process is highly correlated very near to $V_b$. They 
defined a new quantity named as breakdown 
susceptibility\index{breakdown susceptibility} $\chi~(=dn/dV)$. 
Here $n(V)$ is the average number of broken bonds at an external voltage
 $V~(<V_b)$ for a fixed value of $p$. It is obvious that for a sufficient 
voltage $V$, $n(V)$ saturates to $L^d$. $\chi$ exhibits a maxima at 
$V_b^{eff}$  which is different from $V_b^{fin}$, 
where $V_b^{fin}$ is the voltage required to creat the last member of 
the sample spanning conducting cluster. $V_b^{eff}$ approaches $V_b^{fin}$ 
with increasing sample size $L$. It seems that $\chi$ is a divergent
quantity for infinite sample. So there exists a possibility to predict 
$V_b^{fin}$ without going to the complete breakdown point of the sample.
Statistics of the growing clusters upto $V_b$ (minimum external 
voltage required to have the global connection across the sample 
via the conducting sites) in $2d$ site percolation
 was studied by Acharyya \textit{et al.} \cite{Acharyya et al:1996_SCR}. 
$V_b$ is identified not 
only as the point of global breakdown but also the point of divergence
of the rate of the various statistical quantities like total number 
of conducting sites, the average size of the conducting cluster and the 
number of such clusters.     																																																																									

%% file: qua2.tex
\
\label{sec:qua_SCR}
From the microscopic point of view, current conduction requires
the mobility of the electrons across the material and for composites
of metals and dielectrics or insulators this means that there must be 
percolating paths of conducting material in the system  
(as long as electron is considered a classical particle). 
If $p$ denotes the 
concentration of the conducting material in the composite and $p_c$ 
the percolation threshold\index{threshold!percolation}, 
then from our previous discussion we know that 
the conductivity will go to zero at $p_c$ as $|p-p_c|^{t_c}$  
\cite{Stauffer and Aharony:1992_SCR}. For $p$ below $p_c$, the composite is 
ceases to be conducting and the classical breakdown voltage needed to 
make the system conducting goes as $|p-p_c|^{\phi_b}$ 
(see sec.~\ref{sec:unify_SCR}).

Studies on Anderson transition\index{Anderson transition} 
(see ref.~\cite{Lee and Ramakrishnan:1985_SCR} 
for 
example) show that electron as a quantum particle cannot diffuse even through
the geometrically percolating path
due to the coherent back scattering (of the wave function) from 
the random geometry of the cluster in a dimension less than three. Since 
all the states on any percolating lattice gets localized (exponentially) 
electrons do not diffuse through the disordered (percolating) lattice.

In a three dimensional disordered systems there exist a mobility 
edge\index{mobility edge} (a sharp energy 
level in conduction band) $\epsilon_c$ below which all the states are 
localized and above which all the states are extended. 
Insulting or conducting phase appears depending upon the position of  Fermi 
level $\epsilon_f$ below or above the mobility edge $\epsilon_c$. The 
(Anderson) transition\index{Anderson transition} from the insulating 
phase $(\epsilon_f < \epsilon_c)$ to 
metallic phase $(\epsilon_f > \epsilon_c)$ (where the electrons are quantum 
mechanically percolating that is their wave functions extend all over) across 
the mobility edge\index{mobility edge} in a percolating disordered solid, 
has already been well 
studied \cite{Lee and Ramakrishnan:1985_SCR}. In metallic phase the conductivity
increases as $|\epsilon_f - \epsilon_c|^{t_q}$; $t_q \not= t_c$.

In the case of a classical nonpercolating $(p < p_c)$ system, 
an application of an external electric field $E_b$ forces the system to 
undergo breakdown following the relation $E_b \sim |p_c - p|^{\nu}$ 
(see sec.~\ref{sec:critical_SCR}).
In analogy with the above classical dielectric breakdown problem 
one can think of a problem of quantum dielectric 
breakdown\index{dielectric breakdown!quantum version} (possibility of 
the appearance of a conducting path) in 
Anderson insulating ($\epsilon_f < \epsilon_c$ or quantum mechanically 
non-percolating) phase with an application of strong electric field. 
This may be 
thought as a kind of complementary\index{Zener breakdown!quantum version} 
problem of the Zener breakdown\index{Zener breakdown} in 
insulator.

In an insulator, states are localised exponentially within 
the atomic distance $a$ and effective band width $w$ is in general larger 
than $a$. There exists a possibility of tunneling of the states across the 
band gap if the energy gained by an electron in travelling an atomic 
distance $a$ in an electric field $E$ is larger or equal to the band 
gap $\triangle \epsilon_b$ (or when $eEa\geq\triangle \epsilon_b$).  Here $e$ 
denotes the electronic charge and the band gap is the separation between 
conduction and valance band.
In Zener breakdown\index{Zener breakdown}, bands get effectively 
tilted in strong electric 
field $E$ in 
the direction of the field, reducing the band gap effectively  
from $\triangle \epsilon_b$ to $\triangle \epsilon_{br}$. If the effective 
width of the reduced band gap $(w=\triangle \epsilon_{br} / eE)$ becomes 
of the order of atomic distance $a$ interband tunnelling 
takes place for $w \leq a$ and the insulation breaks down. So the  
Zener breakdown\index{Zener breakdown} field $E_b$ scales linearly 
with the band gap: $E_b = \triangle \epsilon_{b} / ea$.

Similar\index{Zener breakdown!quantum version} kind of breakdown of insulation can occur in the case of 
Anderson insulators  with 
an application of strong electric field $E$ in more than 
two dimension. States are localized (exponentially) with in the localization 
length $\xi_q$, where $\xi_q$ varies 
as $\xi_q \sim |\triangle \epsilon_m|^{-\nu_q}$. Here $\triangle \epsilon_m$ 
is the mobility gap; $\triangle \epsilon_m \equiv \epsilon_c - \epsilon_f$.
One can now easily think of a possibility of tunnelling across 
the mobility gap, if 
the energy gained by an electron in travelling
a distance $\xi_q$ in field $E$ is of the order of the 
mobility gap $\triangle \epsilon_m$ (or 
when $e E \xi_q \geq \triangle \epsilon_m$). Thus in contrary to 
the standered Zener breakdown in semiconductors, 
here breakdown field $E_b$ scales 
almost quadratically with mobility gap as \cite{Chakrabarti:1994_SCR} 
\begin{equation}
E_b = \frac{\triangle \epsilon_m}{e \xi_q} \simeq |\triangle \epsilon_m|^{T_q};
\hspace{1mm}T_q=1+\nu_q
\end{equation} 
($\nu_q \sim 0.9$ \cite{Lee and Ramakrishnan:1985_SCR,Shapiro:1983_SCR}, giving 
$T_q \sim 1.9$ in $3d$).

\begin{figure}[htbp]
   \begin{center}
      \includegraphics[width=1.00\textwidth]{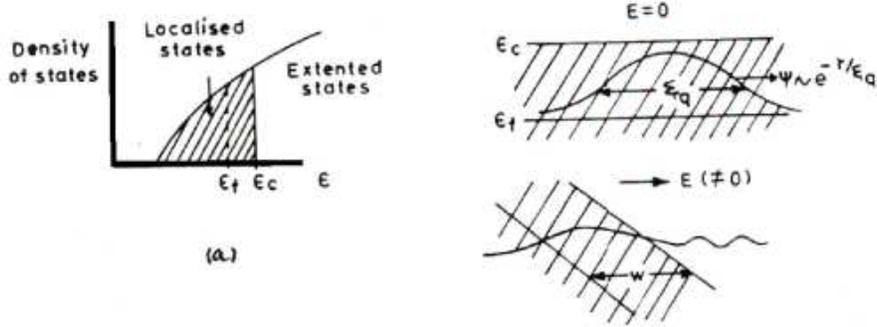}\hspace{2mm}
       \caption{\footnotesize{Schematic density of states for Anderson 
        insulators ($d>2$) shown in (a), where the Fermi label $\epsilon_f$
        is below the mobility edge $\epsilon_c$ (metallic phase for 
        $\epsilon_f>\epsilon_c $). For strong electric field $E$, the band 
        of (localized) states get tilted and tunnelling occurs when the 
        effective width $w (=|\triangle\epsilon_{mr}|/eE)$ of the mobility gap
        $\triangle\epsilon_{mr}$, is less than or equal to the localization
        length $\xi_q$.}}
      \label{fig:quantum_SCR}
   \end{center}
\end{figure}

The tunnelling probability $g(E)$ decreases exponentially with the width of 
the effective barrier as $g(E) \sim exp(-w/\xi_q) \sim exp(-\triangle 
\epsilon_m/eE\xi_q )$ $ \sim exp(-|\triangle \epsilon_m|^{T_q}/E)$.
The cumulative failure probability of Gumbel form (same as for 
fracture \cite{Duxbury and Leath:1987_SCR,Duxbury et al:1986_SCR,Duxbury et al:1987_SCR}
 ) for a sample of size $L$ under field
$E$ is given by \cite{Chakrabarti:1994_SCR}

\begin{eqnarray}
F_L(E) & \sim & 1- exp[-L^d g(E)]
\nonumber
\\
     & \sim & 1-exp\left[-L^d exp \left(-\frac{|\triangle \epsilon_m|^{T_q}}{E}\right)\right]. 
\end{eqnarray}
     
This gives 

\begin{equation}
E_b \sim \frac{|\triangle \epsilon_m|^{T_q}}{\ln L},
\end{equation}
as the size dependence of the typical breakdown field for and above which 
$F_L(E)$ is significant.


%% file: conclu.tex

We have discussed the classical failure of the fuse systems, the dielectric 
breakdown\index{dielectric breakdown} and the quantum 
breakdown\index{dielectric breakdown!quantum version} 
in the Anderson insulators. 
We have discussed how the extreme value statistics and the resulting Gumbel 
distribution\index{distribution function!Gumbel} arises in breakdown 
and failure processes, especially when the 
disorder concentration is low. At high concentration of disorder near the 
percolation  threshold\index{threshold!percolation}, we have discussed 
how the cross-over might take 
place from extreme value to percolation statistics.
We have discussed the system size dependence that arises at the distribution of 
the failure current at low disorder regime. Finally, the extension of 
Zener breakdown\index{Zener breakdown} phenomenon for band insulators to 
the disorder-induced
Anderson insulators has been discussed in sec.~\ref{sec:qua_SCR}.